\documentclass[twocolumn,aps,superscriptaddress]{revtex4-2}
\usepackage[utf8]{inputenc}
\usepackage{amsmath}
\usepackage{amsfonts}
\usepackage{graphicx}
\usepackage{hyperref}

\begin{document}

\title{Compressed Sensing for Efficient Fidelity Estimation of GHZ States}

\author{Farrokh Labib}
\email{farrokh@unitary.foundation}
\affiliation{Unitary Foundation}
\author{David Nicholaeff}
\affiliation{New Mexico Consortium}
\author{Vincent Russo}
\affiliation{Unitary Foundation}
\author{William J. Zeng}
\affiliation{Unitary Foundation}
\affiliation{Quantonation}

\date{\today}

\begin{abstract}
     Accurately characterizing multipartite entangled states is a critical
     challenge in quantum information processing. In this work, we focus on
     applying compressed sensing techniques to efficiently estimate the fidelity
     of Greenberger-Horne-Zeilinger (GHZ) states. By exploiting the inherent
     sparsity of these states, our compressed sensing protocol drastically
     reduces the measurement overhead traditionally required for state
     verification while maintaining high accuracy. To evaluate the practical
     performance of this approach, we test the protocol on GHZ states using both
     quantum simulators and Quantinuum's trapped-ion hardware. Furthermore, we
     implement error detection techniques during our hardware evaluations,
     demonstrating the robustness and viability of compressed sensing for
     fidelity estimation in noisy experimental environments.
\end{abstract}

\maketitle

\section{Introduction}
 The creation and verification of large-scale entangled states are fundamental
 tasks in quantum computing technology. Among these, the
 Greenberger-Horne-Zeilinger (GHZ) state\cite{greenberger1989going}—a maximally
 entangled state of $N$ qubits—serves as a critical benchmark for quantum
 hardware performance, vital for applications in quantum secret
 sharing~\cite{hillery1999quantum}, quantum
 metrology~\cite{giovannetti2004quantum}, error
 correction~\cite{shor1995scheme}, and many others.
    
To utilize GHZ states as benchmarks, we have to reliably and efficiently
estimate the fidelity of the prepared quantum state. Standard quantum state
tomography, or Direct Fidelity Estimation~\cite{flammia2011direct,
dasilva2011practical}, is prohibitively expensive for large $N$, and while
Parity Oscillation methods~\cite{monz2011entanglement, song2017entanglement,
guhne2007toolbox} are more efficient, they still require a large number of data
points to satisfy the Nyquist-Shannon sampling theorem. To overcome these
limitations, we observe that parity oscillation signals from the quantum
measurements are ``sparse signals''. This allows us to use tools from the field
of compressed sensing (CS)~\cite{candes2006near} to exponentially reduce the
number of data points needed to reconstruct the parity oscillation signal and
hence the coherence of the prepared GHZ state. We want to note that compressed
sensing as a general technique has already been widely used in quantum
information, for example, in state tomography~\cite{gross2010quantum,
barreto2026compressed}, Hamiltonian learning~\cite{shabani2011estimation,
ma2024learning}, and there are many other examples.
    
To evaluate the practical performance of this compressed sensing approach, we
test the protocol on GHZ states using both quantum simulators and actual quantum
hardware. However, preparing large GHZ states remains a significant challenge
due to their extreme sensitivity to imperfections in the experiment. There have
been many attempts at preparing large-scale GHZ states in many different quantum
computing platforms like ion-traps~\cite{sackett2000experimental,
leibfried2005creation, monz2011entanglement}, neutral
atoms~\cite{omran2019generation, song2019generation},
photons~\cite{wang2018entanglement}, and
superconducting~\cite{song2017entanglement, wei2020verifying,
mooney2021generation, bao2024creating, liao2025achieving}. The most recent and
largest-scale GHZ state is due to the recent work by
IBM~\cite{javadiabhari2025big}, where they prepare a 120-qubit GHZ state with
fidelity $>0.5$. The main tools used in that work are based on the low-overhead
error detection ideas developed in~\cite{martiel2025low} by the same group. The
main idea is to use pairs of qubits in the GHZ states to apply a parity check
that maximizes ``coverage'' (this is defined in
Section~\ref{sec:error_detection}).
    
Building upon these recent low-overhead error detection strategies, our work
adapts this framework to the all-to-all connectivity of trapped-ion systems such
as Quantinuum hardware. First, this allows us to prepare GHZ states of size $N$
in $\log(N)$ depth, significantly reducing errors. Second, the choice of parity
checks becomes significantly more simplified, and we can reach much higher
coverage of qubits with fewer parity checks, thus reducing the amount of extra
operations needed and thus reducing the potential of adding more noise. 
    
The aim of this paper is twofold: 1) we want to show that there are compressed
sensing techniques that can be used to efficiently estimate the fidelity of GHZ
states using the parity oscillation method, and 2) that this protocol can be
successfully tested on simulators and actual hardware by applying an
error-detection protocol using flag qubits, which works particularly well on
all-to-all connected devices.
	
\section{Preliminaries}
    
In this section, we lay the groundwork for the methods we use. We begin by
describing the structure of GHZ states and the specific error detection strategy
using flag qubits. We then outline the theoretical basis for efficient
verification of the fidelity: first, by detailing the parity oscillation method
for measuring coherence, and second, by introducing the compressed sensing
framework that allows us to estimate fidelity efficiently. \subsection{Error
detection on GHZ states}\label{sec:error_detection} The $N$-qubit GHZ state is a
maximally entangled state defined as
\begin{equation}
    |\text{GHZ}\rangle = \frac{1}{\sqrt{2}}(|0\rangle^{\otimes N} + |1\rangle^{\otimes N})
\end{equation}
It has its stabilizer group generated by $Z_iZ_{i+1}$ for $i=1,\dots,n-1$ and
$XX\dots X$. Therefore, we can coherently check the parity of any two qubits
$i,j$ by introducing an ancilla qubit and applying CNOTs with control qubits $i$
and $j$ to the ancilla. Post-selecting on the ancillas being in the zero state
gives us an error detection protocol. We call these ancilla qubits flag qubits.
	
In this work, the $N$-qubit GHZ state is prepared using a binary tree structure
of CNOT gates, starting from a root Bell pair and expanding outwards. This
logarithmic-depth circuit is efficient but susceptible to errors near the root,
which can propagate to affect a large fraction of the system (high-weight
errors). Error detection protocols employ flag qubits to detect dangerous faults
that could otherwise propagate and corrupt the logical state. In our approach,
we leverage the flexibility of the assumed all-to-all connectivity to place
these flag qubits strategically within the GHZ state preparation circuit.
	
To enhance fidelity using error detection and post-selection, we implement a
flag strategy utilizing optimized $ZZ$ checks: we employ flag qubits to perform
$ZZ$ measurements on subsets of the data qubits at the end of the preparation
circuit such that we maximize the ``coverage'' of that
check~\cite{javadiabhari2025big}.
	
\begin{figure*}
    \centering
	\includegraphics[width=0.5\textwidth]{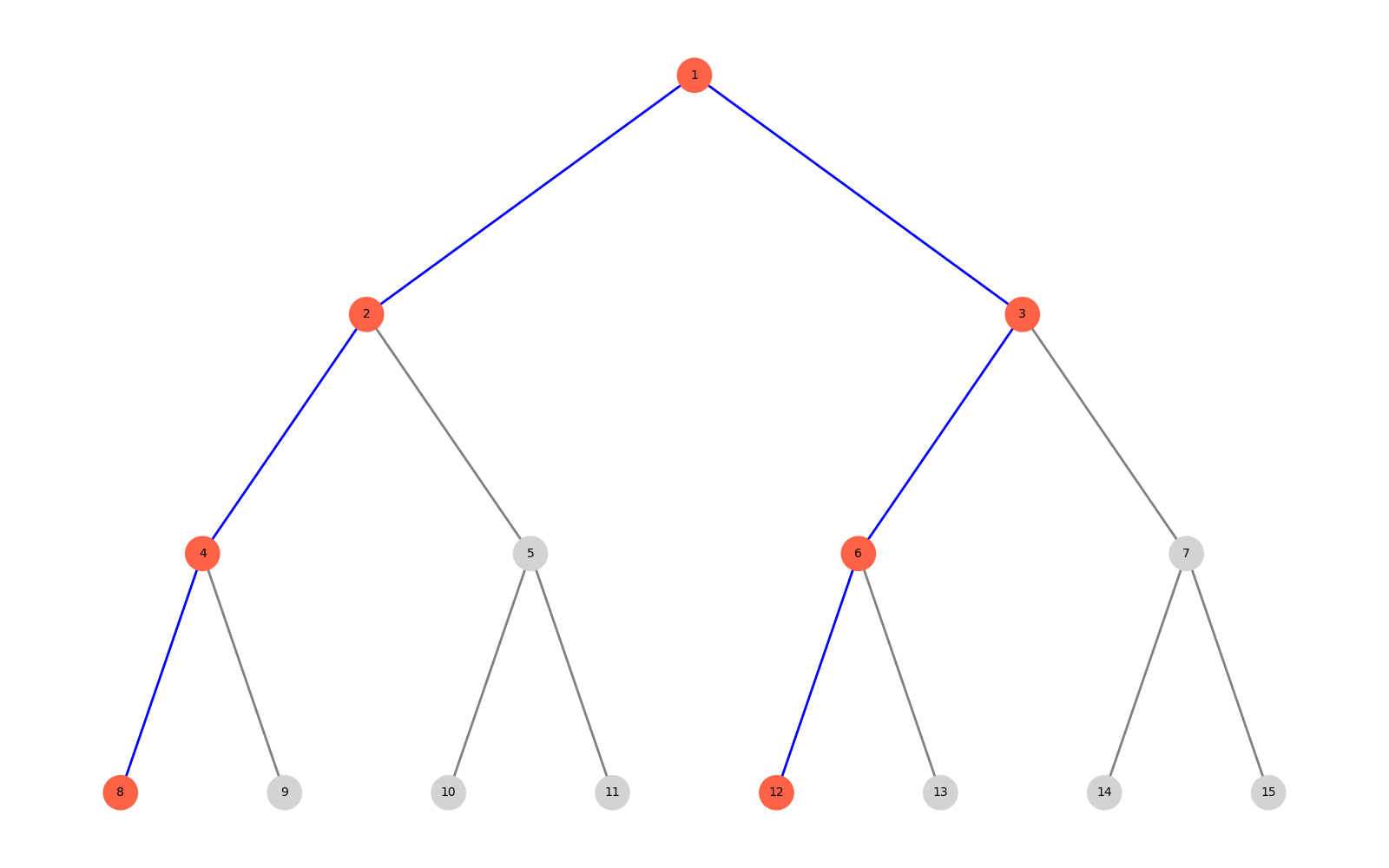}
	   \caption{Binary tree of depth four. The two red leaf nodes are the qubits we
	   use for a parity check, and the paths to their least common ancestor
	   (following the CNOTs) are highlighted in red too. In this case, using just
	   one parity check covers $46.67\%$ of the qubits.}
	   \label{fig:binary_tree}
\end{figure*}

The coverage of a check is defined to be the size of the path from the two
qubits on which the check is applied to the least common ancestor following the
path of the CNOTs backwards in time. By the size of the path, we mean the number
of qubits we visit on this path. The coverage ratio is this size divided by the
total number of qubits. Intuitively, for a given check, the larger the coverage,
the more errors it can potentially detect. Any bitflip error that might happen
on a qubit in the path of a check will propagate to the flag qubit. So we want
to pick pairs of qubits in the GHZ state such that we maximize the total
coverage of all the flag qubit checks. Figure~\ref{fig:binary_tree} shows the
CNOT structure of a logarithmic depth GHZ preparation circuit with an example of
a selected pair of qubits together with the coverage.
Using a greedy algorithm, we identify the optimal pairs of qubits to connect via
a flag to maximize this coverage. This ensures that a small number of flags can
effectively monitor the most critical error pathways in the large entangled
state.

We note that this flag placement optimization can be viewed as an instance of
the classical \emph{maximum coverage problem}: given a collection of sets---each
defined by a pair of leaf qubits and the nodes along their respective paths to
the least common ancestor in the preparation tree---select $k$ sets whose union
is maximized. The objective function $f(S) = |\bigcup_{(i,j) \in S}
\mathrm{path}(i,j)|$ is monotone submodular, since the marginal gain from adding
a new flag pair diminishes as earlier pairs already cover overlapping portions
of the tree. A classical result of Nemhauser, Wolsey, and
Fisher~\cite{nemhauser1978analysis} guarantees that the greedy algorithm
achieves at least a $(1 - 1/e) \approx 0.632$ fraction of the optimal coverage,
and this ratio is known to be the best achievable in polynomial time under
standard complexity-theoretic assumptions~\cite{feige1998threshold}. This
provides a formal worst-case performance guarantee for the greedy flag placement
strategy we employ.

Finally, we utilize a post-selection scheme: any experimental shot where a flag
qubit signals an error (measures `1') is discarded. This yields a ``heralded''
state with significantly higher fidelity as we will show empirically using noisy
simulators and Quantinuum hardware in Sections~\ref{sec:simulator_results}
and~\ref{sec:hardware_results} respectively.

\subsection{Estimating fidelity using parity oscillation}

The parity oscillation method~\cite{monz2011entanglement, song2017entanglement,
guhne2007toolbox} is a standard approach for estimating state fidelity by
decomposing the GHZ state into its population and coherence components. The
fidelity $F$ is defined as
\begin{equation}
    F = \langle \text{GHZ} | \rho | \text{GHZ} \rangle = \frac{1}{2}(\langle P \rangle + \langle \chi \rangle)
\end{equation}
where $\langle P \rangle = \langle 0 | \rho | 0 \rangle + \langle 1 | \rho | 1 \rangle$ represents the population, which is estimated directly from computational basis measurements.
	
The coherence $\langle \chi \rangle = \langle 0 | \rho | 1 \rangle + \langle 1 |
\rho | 0 \rangle$ requires measurements in different bases. This is achieved by
measuring the global parity operator $\mathcal{P}(\phi)$ after applying a
rotation $R_z(-\phi)$ followed by $R_y(-\pi/2)$ to all $N$ qubits
\begin{equation}
    \mathcal{P}(\phi) = \langle \bigotimes_{j=1}^N (\cos\phi X_j + \sin\phi Y_j) \rangle
\end{equation}
For an $N$-qubit GHZ state, this signal oscillates as $C \cos(N\phi + \theta)$.
The phase-offset $\theta$ is the result of coherent noise acting on the qubits
and results in the rotated GHZ state
\begin{equation}
    |\text{GHZ}_\theta\rangle = \frac{1}{\sqrt{2}}(|0\rangle^{\otimes N} + e^{i\theta}|1\rangle^{\otimes N}).
\end{equation}
To estimate the coherence $C$ and phase-offset $\theta$, one may perform a
non-linear least squares curve fit to the parity data collected at various
angles $\phi$. Alternatively, a Fourier transform can be applied to extract the
signal components
\begin{equation}
    I_q = \frac{1}{2(N+1)} \sum_{j=0}^{2N+1} e^{\frac{iqj\pi}{N+1}} \langle \mathcal{P}(\frac{j\pi}{N+1}) \rangle
\end{equation}
In this framework, the absolute coherence is determined by $C = |I_N| +
|I_{-N}|$. Both methods are physically equivalent and provide a robust estimate
of the state's off-diagonal elements. The state is certified as having genuine
multi-partite entanglement if the estimated fidelity $F > 0.5$. With a non-zero
phase-offset $\theta$, the estimated fidelity with respect to the ideal GHZ
state will be lower than with respect to the rotated version
$|\text{GHZ}_\theta\rangle$. Both have genuine multi-partite entanglement.
    
To satisfy the Nyquist-Shannon sampling theorem and avoid aliasing, one has to
measure the parity signal at $2N$ equally spaced angles $\phi =
\frac{k\pi}{N+1}$ for $k=0,1,\dots, 2N-1$.

\subsection{Compressed sensing for sparse signals}
Here we follow the reference~\cite{foucart2013mathematical} on compressed
sensing, but only focus on the parts necessary for our application.

The compressed sensing problem consists in reconstructing a sparse vector $x\in
\mathbb{C}^N$ from
\begin{equation}
    y=Ax,
\end{equation}
where $A\in\mathbb{C}^{M\times N}$ is the so-called measurement matrix and $y$
the measurement values. The vector $x$ is the coefficient vector of the signal
$f$ in some orthonormal basis $\{\psi_k\}_k$,
\begin{equation}
    f(t) = \sum_kx_k\psi_k(t).
\end{equation}
Here $M$ is the number of measurement values and we wish to minimize this while
still be able to recover $x$. The efficiency in compressed sensing comes from
the randomness of the measurement matrix $A$. Let $t_1, \dots, t_M$ be random
sampling points and suppose we are given the values 
\begin{equation}
    y_l = f(t_l) = \sum_kx_k\psi_k(t_l).
\end{equation}
In this case, the measurement matrix $A$ has entries $A_{l,k} = \psi_k(t_l)$.
The task of compressed sensing is to reconstruct $f$ from the samples vector $y$
and to perform this task with as few samples $m$ as possible. We will be mainly
interested in the orthonormal basis consisting of $\{\sin(kt), \cos(kt)\}_k$,
which is also known as the Discrete Cosine Transform (DCT). 
    
In the context of GHZ verification, the parity oscillation signal
$\mathcal{P}(\phi)$ can be expanded as
\begin{equation}
    \mathcal{P}(\phi) = C \cos(N\phi + \theta) = C \cos\theta \cos(N\phi) - C \sin\theta \sin(N\phi)
\end{equation}
We can view this as a DCT of a signal that is extremely sparse, possessing
non-zero coefficients only at frequency $N$. We construct the measurement matrix
$A$ of candidate frequencies $k \in \{1, \dots, N\}$, where $N$ is the system
size of the GHZ state. The measurement matrix contains columns for both cosine
and sine components
\begin{equation}
    A_{k,i} = (\cos(k\phi_i), -\sin(k\phi_i))
\end{equation}
We collect data $y_i = \mathcal{P}(\phi_i)$ at $M$ randomly chosen angles
$\phi_i$, where $M \ll 2N$. The recovery problem is formulated as an
$L_1$-regularized least squares optimization (Lasso)
\begin{equation}
    \min_{x} \frac{1}{2M} ||y - Ax||_2^2 + \alpha ||x||_1
\end{equation}
The $L_1$ penalty promotes sparsity, forcing most coefficients in $x$ to zero
and isolating the true frequency component $n_{rec}$.

However, empirically, we saw that Lasso estimation introduces a bias (shrinkage)
to the coefficients. To correct this, we perform a two-step procedure
\begin{enumerate}
    \item \textbf{Support detection}: Use Lasso to identify the dominant
    frequency $n_{rec}$ by selecting the index $k$ with the largest coefficient
    magnitude $\sqrt{a_k^2 + b_k^2}$.
    \item \textbf{Parameter refinement}: Perform an unregularized Ordinary Least
    Squares (OLS) fit using only the columns corresponding to $n_{rec}$. This
    yields unbiased estimates for $a = C \cos\theta$ and $b = C \sin\theta$.
\end{enumerate}
Finally, the coherence parameters are recovered as $C = \sqrt{a^2 + b^2}$ and
$\theta = \arctan(b/a)$.

The theoretical foundation for this efficiency lies in the Restricted Isometry Property (RIP)~\cite{candes2006near}. For a signal that is $s$-sparse in the DCT basis of size $N$, compressed sensing theory guarantees stable recovery with high probability provided the number of random measurements $M$ satisfies:
\begin{equation}
    M \ge c s\log^2(s) \log(N)
\end{equation}
where $c$ is a constant. In our case, the signal contains only a single frequency component (plus its conjugate), meaning $s=2$. Consequently, the required number of measurements scales logarithmically with the system size, $M \propto \log(N)$. This represents a double exponential reduction in measurement overhead compared to full tomography, and an exponential improvement over standard parity oscillation, which typically requires $\mathcal{O}(N)$ samples. This scaling allows us to efficiently verify even very large entangled states.

Crucially, in realistic experimental settings, the parity measurements $y_i$ are corrupted by statistical shot noise and hardware imperfections, yielding $y = Ax + e$ with bounded noise $\|e\|_2 \le \eta$. The RIP further guarantees that the reconstruction error is bounded by:
\begin{equation}
    \|x - x^*\|_2 \le C_1 s^{-1/2} \|x - x_s\|_1 + C_2 \eta
\end{equation}
Because the ideal GHZ parity signal is exactly $2$-sparse ($s=2$), the first term (compressibility error) vanishes, and the reconstruction error is dominated entirely by the measurement noise $\eta$. Since $\eta$ scales as $1/\sqrt{S}$ for $S$ experimental shots, this provides a formal guarantee that the compressed sensing estimator will stably converge to the true coherence as the shot count increases, without requiring a dense grid of measurement angles. Furthermore, this noise bound provides theoretical justification for our two-step procedure: the support detection via Lasso is stable under bounded noise $\eta$, and once the support is correctly identified, the unregularized OLS fit serves as the optimal unbiased estimator on that support.

We remark that the single-frequency recovery problem we solve has connections to
several well-studied areas in classical computer science and signal processing.
The sparse Fourier transform literature~\cite{hassanieh2012nearly} addresses the
general problem of recovering $K$-sparse frequency representations in sublinear
time; for the $K = 1$ case relevant here, classical spectral estimation methods
such as Prony's method~\cite{stoica2005spectral} can recover the frequency
algebraically, though they are considerably less robust to noise than the
$L_1$-regularization approach we employ. Our measurement strategy also parallels
non-adaptive combinatorial group testing~\cite{du2000combinatorial}, where
pooled tests identify a small number of ``defective'' items in a large
population. Each random-angle parity measurement acts as a pooled test that
mixes all frequency components, and the logarithmic scaling $M \propto \log(N)$
mirrors the information-theoretic lower bounds for identifying a constant number
of active items among $N$ candidates.

\section{Experiments}

To validate our approach, first, we perform numerical simulations to benchmark
the accuracy of the compressed sensing estimator and the efficacy of the flag
qubit protocol under controlled noise models. Second, we verify our findings on
actual quantum hardware, demonstrating high-fidelity GHZ state preparation and
efficient verification on Quantinuum's trapped-ion processor.	
	
\subsection{Verification of compressed sensing estimation}
Before using the compressed sensing protocol on hardware data, we verify its
accuracy and scalability using numerical simulations. The goal is to confirm
that the coherence $C$ estimated from a logarithmic number of random samples ($M
\propto \log N$) reliably tracks the true state coherence, even in the presence
of noise.
	
\subsubsection{Methodology}
We implemented a verification pipeline comparing the Compressed Sensing
(CS)-estimated coherence ($C_{est}$) against an exact theoretical benchmark
($C_{exact}$) for GHZ states of size $N \in [5, 40]$. To handle the simulation
complexity, we used a hybrid strategy:

\begin{itemize}
    \item \textbf{Small scale ($N \le 10$)}: We performed full density-matrix
    simulations using Qiskit's \texttt{AerSimulator} under depolarizing noise.
    $C_{exact}$ was computed directly from the off-diagonal element
    $|\rho_{0...0,1...1}|$ of the density matrix.
    \item \textbf{Large scale ($N > 10$)}: For larger systems where density
    matrix simulation is intractable, we utilized a ``Fast Emulator''. This
    approach analytically predicts the expected coherence $C_{exp}$ by counting
    the number of single-qubit ($N_{1q}$) and two-qubit ($N_{cx}$) gates in the
    specific circuit. Assuming a depolarizing noise channel with rates $p_{1q}$
    and $p_{2q}$, the coherence decays as
    \begin{equation}
        C_{exp} \approx (1 - p_{2q})^{N_{cx}} (1 - p_{1q})^{N_{1q}}
    \end{equation}
    To simulate experimental data, we sample the parity outcomes directly. For a
    given phase $\phi$, the probability of observing even parity is
    $P_{even}(\phi) = \frac{1}{2}(1 + C_{exp} \cos(N\phi))$. The number of even
    outcomes in $S$ shots is then drawn from the binomial distribution $n_{even}
    \sim B(S, P_{even}(\phi))$, yielding a statistical simulation of the
    measurement process without the cost of vector evolution.
\end{itemize}

For each $N$, we performed 100 independent trials. In each trial, the Compressed
Sensing (CS) estimator reconstructed the signal frequency and amplitude from $M
\approx 5 \ln N$ random phase samples.

\begin{figure*}
    \centering
    \includegraphics[width=0.49\textwidth]{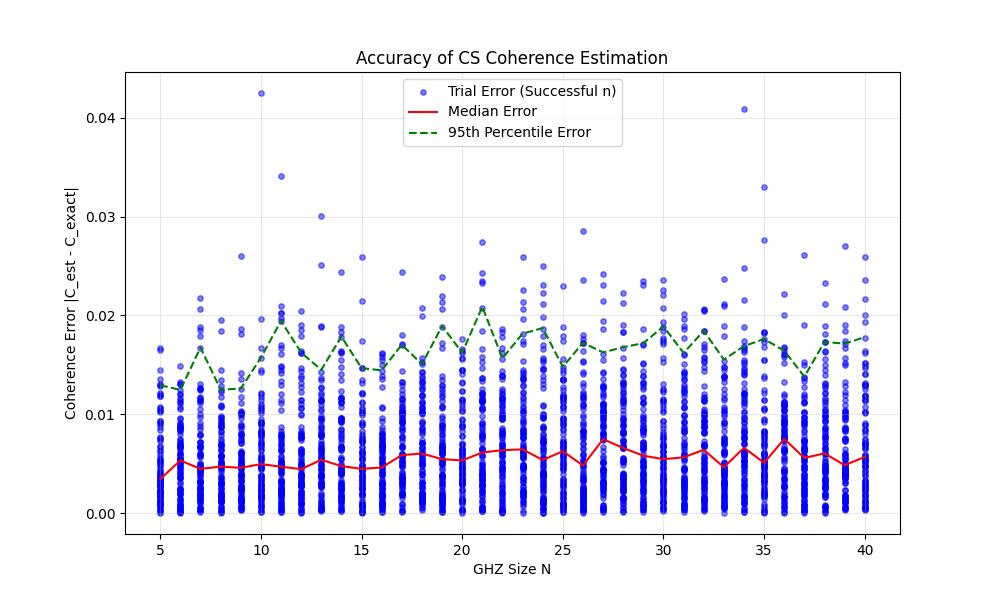}
    \hfill
    \includegraphics[width=0.49\textwidth]{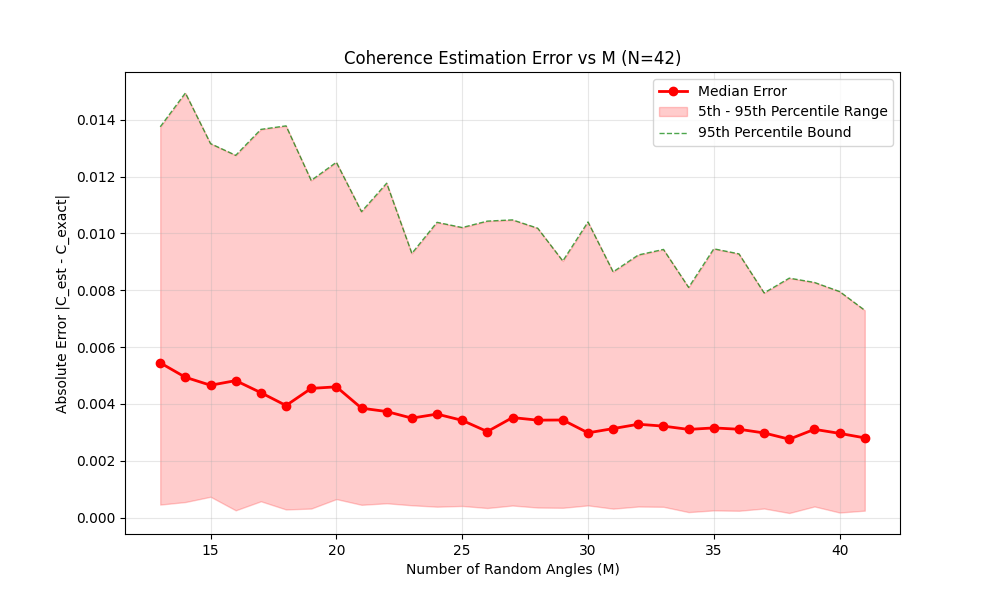}
    \caption{\textbf{Left}: Verification of the compressed sensing estimation.
    The plot shows the median (red line) absolute error $|C_{est} - C_{exact}|$
    and 95th percentile (green dashed line) error, confirming high accuracy
    across all system sizes. The blue points represent individual trial errors.
    \textbf{Right}: Coherence estimation of a fixed-size GHZ state as a function
    of the number of chosen random angles. The plot shows the median absolute
    error $|C_{est} - C_{exact}|$ (red line), the 95th percentile (green dashed
    line), and the range 5th-95th percentile error.}
    \label{fig:cs_coherence_estimation}
\end{figure*}

\subsubsection{Results and measurement bias}
The results, summarized in Fig. \ref{fig:cs_coherence_estimation}, demonstrate
that the CS estimator maintains high accuracy across the entire range of $N$.
The median and 95-percentile error remain bounded, confirming that the
logarithmic sampling scaling is sufficient for reliable coherence estimation.
The depolarizing error rates that we used here are $p_{1q}=0$ and $p_{2q}=0.01$.
We explain now why we used perfect single-qubit gates.

The exact coherence measures the state \textit{before} the parity measurement,
whereas the CS protocol estimates the visibility \textit{after} applying the
necessary rotations ($R_z(\phi)$) and Hadamard gates. These measurement gates
introduce additional single-qubit errors spreading across the $N$ qubits,
scaling as $(1-p_{1q})^N$. Using noisy single-qubit gates would then add noise
to the CS approach, while for the density matrix simulation, that noise would
not be present. This was confirmed by running simulations with noisy and perfect
single-qubit gates, where the discrepancy appeared with noisy single-qubit gates
and disappeared with perfect single-qubit gates. Thus, the CS estimator
correctly reports the fidelity, also accounting for the degradation inherent in
the measurement process itself.

\subsubsection{Frequency identification probability}
The fidelity of the coherence estimation relies on correctly identifying the
oscillation frequency $n$. If the compressed sensing algorithm retrieves the
wrong frequency ($n_{rec} \neq N$), the subsequent amplitude fit will
characterize noise rather than the signal. Also, we wouldn't be certain that the
quantum computer prepared the GHZ state of the correct size: the frequency of
the signal is exactly equal to the size of the GHZ state.

\begin{figure*}
    \centering
    \includegraphics[width=0.65\textwidth]{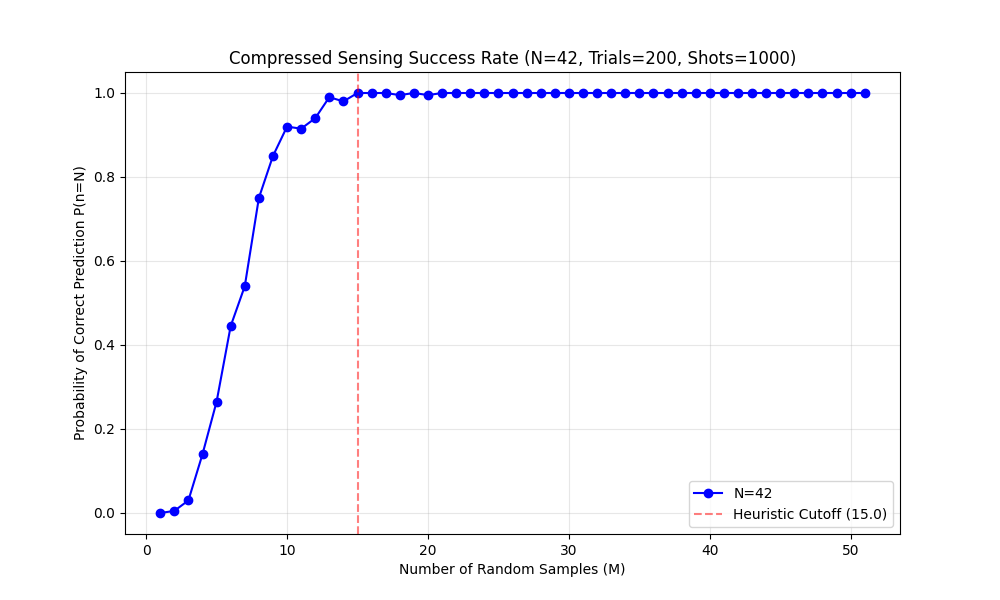}
    \caption{Success probability of recovering the correct frequency component
    $n=N$ as a function of random samples $M$. The vertical line indicates the
    heuristic threshold $15$, beyond which the recovery rate approaches 100\%.}
    \label{fig:cs_success}
\end{figure*}

We define the ``success probability'' as the fraction of trials where the recovered frequency matches the true GHZ size $P(n_{rec} = N)$. Fig. \ref{fig:cs_success} illustrates this probability as a function of the number of random samples $M$ for a GHZ state of size $N=42$. We observe that the success probability rapidly converges as $M$ increases at the start and saturates after $M$ exceeds a heuristic cutoff of $M= 15$ (a bit lower than the $5 \log(N)$ threshold used in Figure~\ref{fig:cs_coherence_estimation}). This empirically confirms the theory that a logarithmic sampling strategy provides sufficient information to uniquely identify the GHZ signal from the sparse spectral domain.

\subsection{Increasing fidelity through parity checks}\label{sec:simulator_results}
To show the effectiveness of the error detection protocol before hardware
execution, we performed numerical simulations using Qiskit's
\texttt{AerSimulator}. We modeled a noisy quantum processor with depolarizing
error channels, setting the single-qubit gate error rate to $p_{1q} = 0.1\%$ and
the two-qubit gate error rate to $p_{2q} = 1.0\%$.

We simulated the preparation and verification of GHZ states for system sizes $N
= 10$ under depolarizing and thermal noise. We varied the number of flag qubit
pairs $k$ incorporated into the circuit, ranging from $k=0$ (standard
preparation without checks) to $k=2$ pairs. The ``coverage ratio''—defined as
the percentage of qubits in the GHZ state whose error paths are monitored by the
flags—increases with $k$.
	
For each configuration, we estimated the state fidelity using our compressed
sensing protocol with $M \approx 5 \ln N$ random measurement angles. The
fidelity was calculated as $F = (P + C)/2$, where the population $P$ was
measured in the $Z$-basis, and the coherence $C$ was recovered from the sparse
parity oscillation signal. To ensure small error bars, we accumulated $30,000$
shots per measurement circuit.

\begin{figure*}
    \centering
    \includegraphics[width=0.6\textwidth]{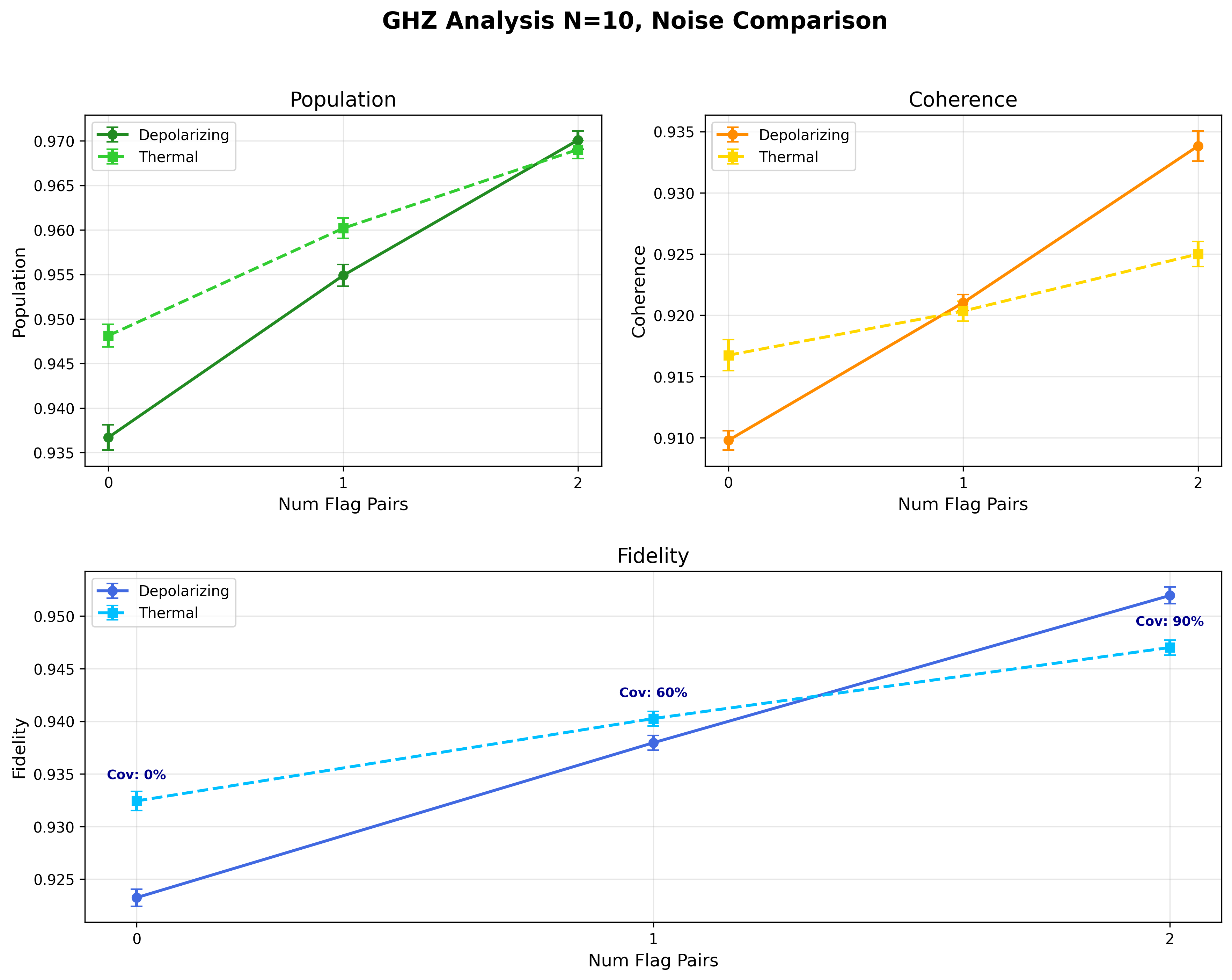}
    \caption{Simulation results for $N=10$ GHZ state preparation under
    depolarizing and thermal noise. The plot shows the estimated fidelity as a
    function of the number of flag qubits used. Annotation labels indicate the
    coverage ratio. We observe a clear monotonic increase in population,
    coherence, and fidelity estimates as more flag qubits are added, confirming
    that the protocol effectively detects and filters out errors propagating
    through the circuit.}
    \label{fig:flag_simulation}
\end{figure*}
	
The results, illustrated in Figs. \ref{fig:flag_simulation} for the $N=10$ case,
demonstrate a clear advantage to using flag qubits under both depolarizing and thermal noise. As the number of flag pairs increases
(corresponding to higher coverage), the post-selected fidelity improves. For
example, moving from 0 flags (0\% coverage) to 2 flags (90\% coverage) yields a
marked increase in fidelity, demonstrating that the flags successfully herald
the presence of errors. This confirms that maximizing the coverage ratio is a
valid proxy for maximizing experimental fidelity.

\subsection{Hardware and emulator results}\label{sec:hardware_results}
\begin{figure*}
    \centering
    \includegraphics[width=0.60\textwidth]{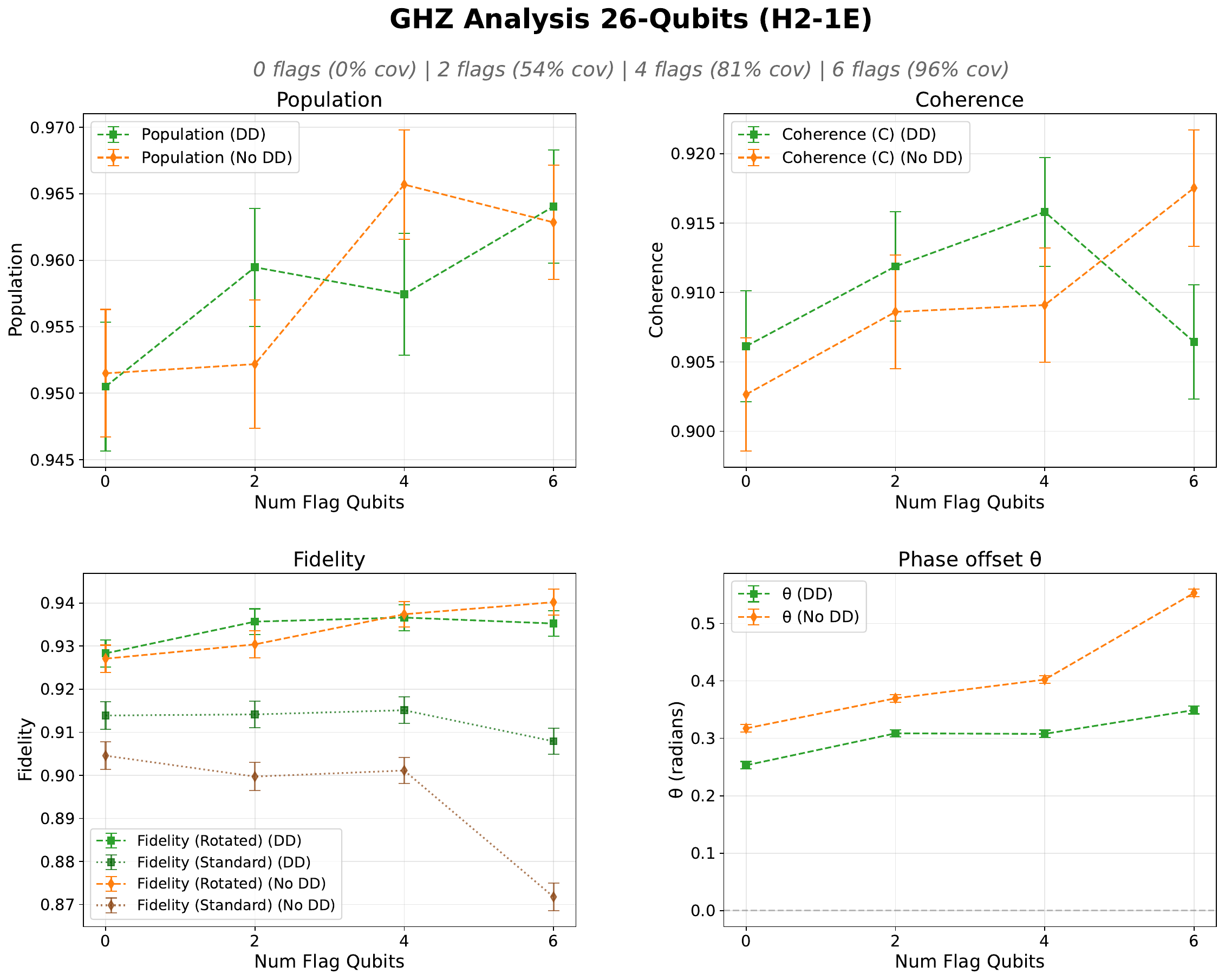}
    \caption{Simulation results for $N=26$ GHZ state preparation using
    Quantinuum H2-1 Emulator with noise. The plot shows the estimated fidelity
    as a function of the number of flag qubits used. We used 20 random angles at
    which we sample. The coverage ratio for several flags can be found at
    the top of the figure. We observe a clear monotonic increase in population
    and rotated fidelity estimates as more flag qubits are added. However, the
    standard fidelity is decreasing for which the reason is the increasing phase
    offset $\theta$, due to unwanted rotations. We have also applied dynamic
    decoupling (DD).}
    \label{fig:flag_simulation_H2E}
\end{figure*}

\begin{figure*}
    \centering
    \includegraphics[width=0.60\textwidth]{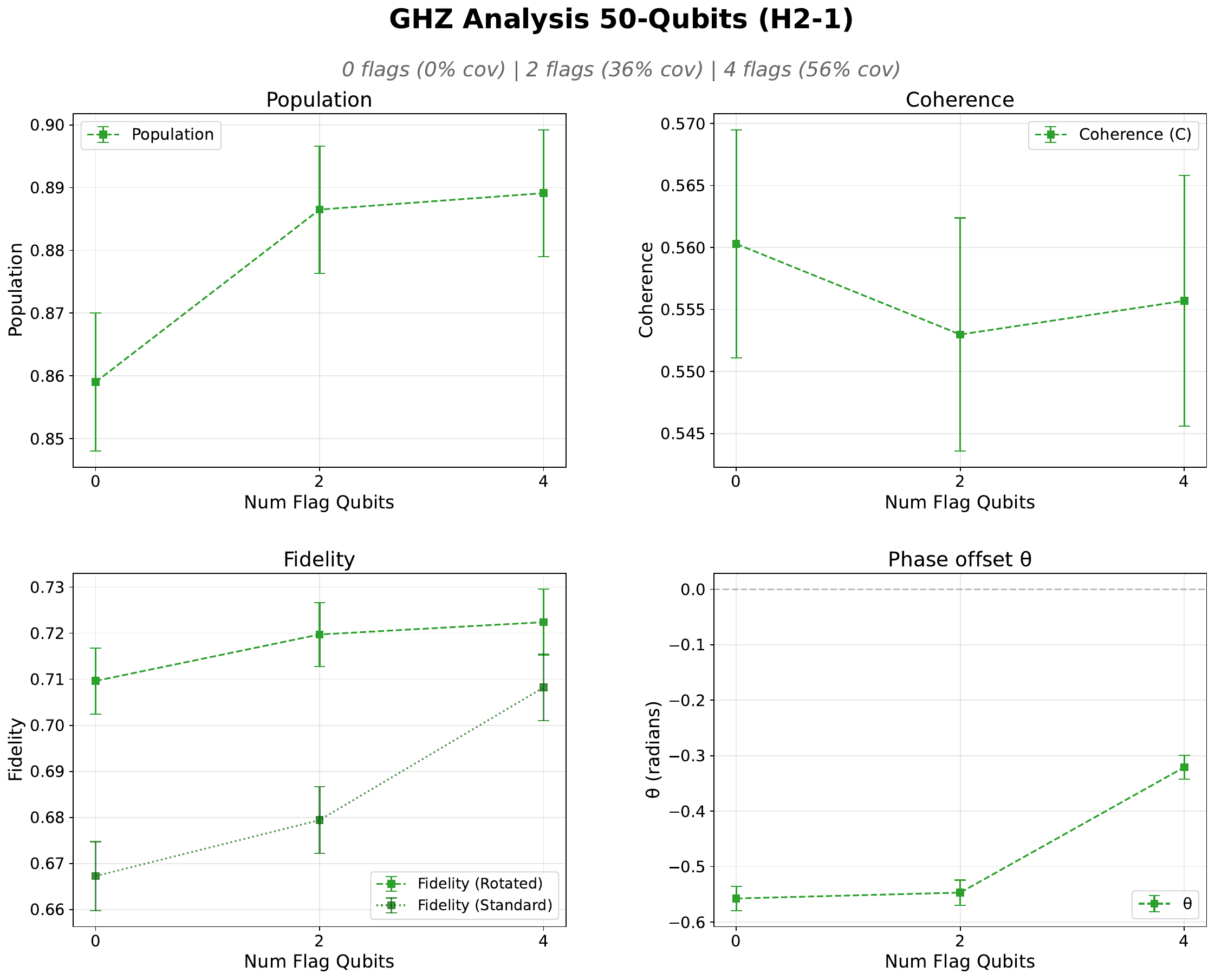}
    \caption{Estimated fidelity of the 50 qubit standard and rotated GHZ state
    on the H2 Quantinuum device. We used 1000 shots on each parity oscillation
    circuit. We used 19 random angles at which we sample. The error bars are
    quite wide as a result of low number of shots to reduce cost. No error
    mitigation has been applied here.}
    \label{fig:hardware_50q}
\end{figure*}

In this section, we show the results of the experiments performed on the
Quantinuum H2-1 quantum computer and on the H2-1E emulator. In Figure~\ref{fig:flag_simulation_H2E} the results are shown for a 26 qubit experiments on the H2-1E emulator. In Figure~\ref{fig:hardware_50q} we show the result of a
50-qubit experiment with a varying number of flag qubits on the $56$-qubit H2-1
device. In both cases one sees that error detection using flag qubits helps in increased fidelity for the rotated GHZ states. 

The divergence between the standard fidelity and rotated fidelity highlights a fundamental trade-off between stochastic and coherent errors. The flag checks effectively detect and filter stochastic bit-flips occurring during state preparation. This post-selection increases the purity and the magnitude of the coherence, thus improving the rotated fidelity. However, executing these extra checks requires additional two-qubit gates and ion transport. In trapped-ion architectures, these extra physical operations introduce coherent over-rotations. For a GHZ state, these local phase errors rapidly accumulate into a global phase offset $\theta$. Because the standard fidelity is penalized by $\cos(\theta)$, this accumulated phase offset suppresses the standard fidelity even as the underlying coherence magnitude improves.

Furthermore, as seen in Figure~\ref{fig:hardware_50q}, a significant phase offset exists even before any flag qubits are added. This is a classic signature of systematic calibration drift. Because a GHZ state is maximally sensitive to global phase ($\theta = \sum_{i=1}^{N} \epsilon_i$), even microscopic coherent phase errors on individual qubits accumulate into a macroscopic phase offset for $N=50$. This extreme sensitivity to minor calibration errors underscores the necessity of calculating the rotated fidelity, which isolates the true structural entanglement from these correctable, coherent shifts.

\subsubsection{Quantum error mitigation}\label{sec:qem}
While flag qubits detect errors during state preparation, they do not correct
for all noise sources present on the hardware. In particular, readout errors
systematically bias the observed bit-string distributions, and low-frequency
dephasing during idle circuit periods degrades coherence. To address these
residual error channels, we layer two quantum error mitigation (QEM)
techniques---individually and in combination---on top of the flag qubit
protocol: Readout Error Mitigation (REM)~\cite{maciejewski2020mitigation,
bravyi2021mitigating, nation2021scalable} and Dynamical Decoupling
(DD)~\cite{viola1999dynamical, viola1998dynamical, zhang2014protected}.

REM corrects measurement bias by applying the inverse of a tensored single-qubit
confusion matrix to the post-selected probability vector. Because the confusion
matrix factorizes as a tensor product over qubits, the correction scales
efficiently to large systems and is applied entirely in classical
post-processing---the quantum circuits remain unchanged. DD targets a
complementary noise source: during idle time steps where a qubit waits while
two-qubit gates act on other qubits, low-frequency dephasing accumulates.
Inserting identity-equivalent gate sequences ($XX$ pairs) into these idle
windows refocuses the qubit and suppresses decoherence. We use the digital
dynamical decoupling implementation from Mitiq~\cite{larose2022mitiq}, applying
$XX$ sequences to all idle qubit slots before circuit submission. We evaluate
four configurations on 25-qubit GHZ states with $k \in \{0, 2, 4, 6\}$ flag
pairs: (i)~unmitigated, (ii)~REM only, (iii)~DD only, and (iv)~REM+DD.

\begin{figure*}
    \centering
    \includegraphics[width=0.7\textwidth]{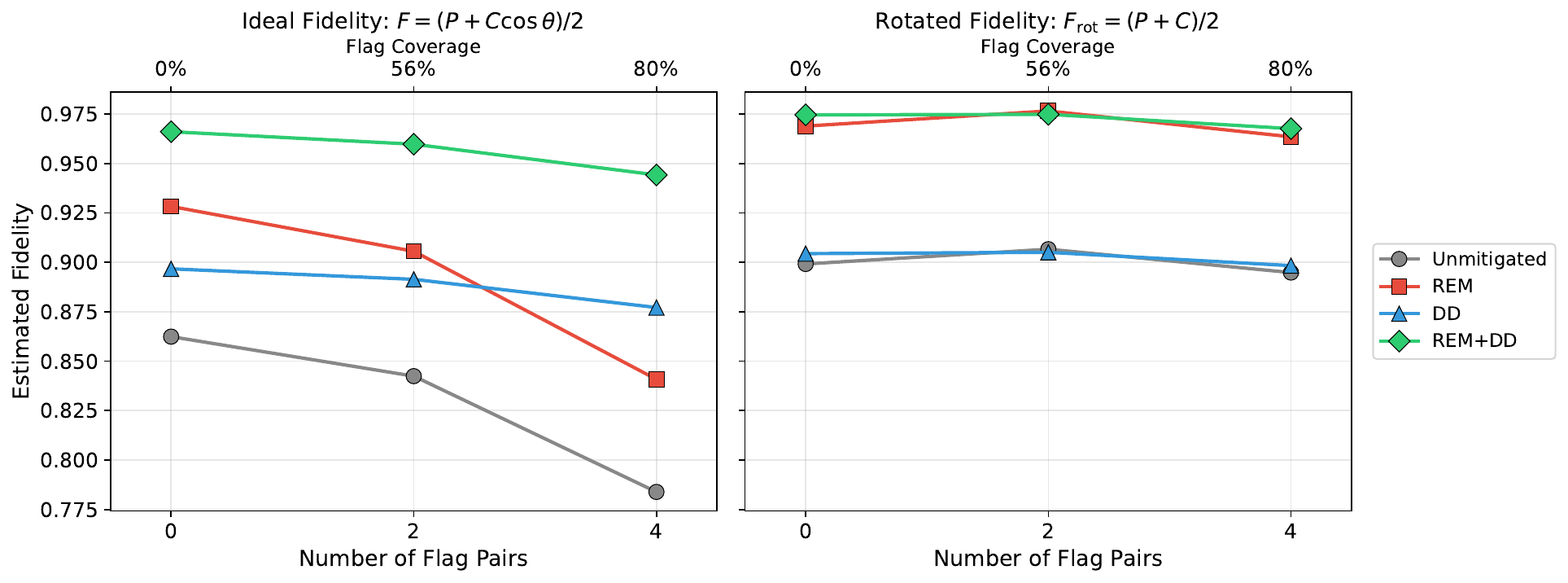}
    \caption{Comparison of QEM techniques on the Quantinuum H2-2E emulator for
    25-qubit GHZ state preparation with compressed sensing verification. The
    estimated fidelity is shown as a function of the number of flag qubit pairs
    ($k \in \{0, 2, 4\}$) for four configurations: unmitigated, REM only
    (analytical correction with $p_{\mathrm{ro}} = 0.2\%$), DD only, and REM+DD.
    REM provides the largest improvement by correcting measurement bias in both
    population and coherence estimates. DD shows a modest effect on the
    emulator, consistent with the emulator's noise model having limited
    idle-time dephasing. The combination REM+DD yields the highest fidelities
    across all configurations.}
    \label{fig:qem_comparison}
\end{figure*}

As shown in Figure~\ref{fig:qem_comparison}, REM provides a massive boost to the estimated fidelity. Because GHZ states are macroscopic, observing the correct population requires all $N$ qubits to be read out perfectly; for $N=25$, even a $0.2\%$ readout error rate per qubit results in approximately a $5\%$ failure rate per shot. By applying the inverse confusion matrix, REM systematically unbiases this large classical error. In contrast, DD provides only a modest improvement. This is likely because the simplified, memoryless Markovian noise models often used in emulators do not fully capture the correlated, low-frequency drifts that DD is explicitly designed to refocus on physical hardware.

Furthermore, unlike the results on the H2-1 hardware and H2-1E emulator, adding flag qubits on the H2-2E emulator fails to improve the overall fidelity. This indicates that the break-even threshold for error detection is not met under the specific noise model calibrated for H2-2E. If the baseline two-qubit gate error rate is very low, mid-circuit faults are rare. In this regime, the additional gates and transport operations required for the parity checks inject more depolarizing and SPAM errors than they successfully catch, resulting in a net decrease in fidelity.

\section{Conclusion}
In this work, we have demonstrated a robust and scalable framework for preparing
and verifying large GHZ states. By combining flag-based error detection with a
compressed sensing verification scheme, we address two critical challenges in
quantum computing technology: the sensitivity of large entangled states to noise
and the prohibitive cost of full state tomography. Our results confirm that
strategically placed flag qubits can enhance state fidelity by heralding errors,
while compressed sensing allows us to verify these states with logarithmic
measurement overhead. This approach offers a practical pathway for benchmarking
GHZ state preparation circuits on next-generation quantum processors.

Our work also reveals connections between quantum state verification and
classical algorithmic theory that may guide future improvements. The flag
placement optimization is a submodular maximization problem on tree-structured
sets, and the binary tree topology of the preparation circuit may admit more
efficient exact algorithms---for example, via Euler tour decompositions or tree
dynamic programming---that go beyond the $(1-1/e)$ approximation guarantee of
the greedy approach. On the verification side, the single-frequency structure of
the parity oscillation signal connects our method to a broader landscape of
spectral estimation techniques and group testing strategies, suggesting
alternative recovery algorithms with different noise-robustness tradeoffs.
Exploring these classical-quantum algorithmic connections is a promising
direction for further improving scalable entanglement verification.

\bibliographystyle{plain}
\bibliography{references}
	
\end{document}